\documentstyle[twoside, epsfig]{article}
\input ibvs2.sty
\begin{document}

\IBVShead{xxxx}{xx xxxxxxx 2010}

\IBVStitletl{Time-resolved spectroscopy of the polar}{RBS~0324($=$1RXS J023052.9$-$684203)\footnote{Based on observations collected at the Southern Astrophysical Research Telescope (SOAR), Cerro Pach\'{o}n, Chile.}}

\IBVSauth{Cieslinski, D.$^{1}$; Rodrigues, C. V.$^{1}$; Silva, K. M. G.$^{1}$ ; Diaz, M. P.$^{2}$}

\IBVSinst{Instituto Nacional de Pesquisas Espaciais, Divis\~ao de Astrof\'isica, Brazil} 

\IBVSinst{Instituto de Astronomia, Geof\'isica e Ci\^encias Atmosf\'ericas, Universidade de S\~ao Paulo, Brazil} 

\SIMBADobj{RBS~0324, 1RXS J023052.9$-$684203}
\GCVSobj{}

\IBVStyp{Cataclysmic variables, polars}
\IBVSkey{Spectroscopy} 

\IBVSabs{We present time-resolved spectroscopy of the polar}
\IBVSabs{RBS~0324($=$1RXS~J023052.9$-$684203) spanning more}
\IBVSabs{than one orbital cycle.}
\IBVSabs{The emission lines present complex and variable profiles.}
\IBVSabs{Radial velocities were determined for the HeII~$\lambda$4686}
\IBVSabs{and H$\beta$ emission lines.}
\IBVSabs{The curves are similar for both lines, with a semi-amplitude}
\IBVSabs{of about 250 km/s.}
\IBVSabs{The modulation of 1.5~mag of the continuum flux is compatible}
\IBVSabs{with those previously presented in the literature.}
\IBVSabs{All these characteristics are typical of polars.}

\begintext
AM~Herculis stars or polars are the cataclysmic variables whose white dwarf presents the strongest superficial magnetic field. The intensity of this field varies from $\sim$10 to 200~MG and is enough to prevent the formation of an accretion disc (with the material coming from the secondary star through L$_{1}$) around the white dwarf and to lead the flux through the magnetic lines. A shock occurs near the white dwarf, resulting in an increase of density and temperature of the gas. This region emits strong X-rays as well as polarised radiation in the visible region, consequently, such characteristics are used to identify objects of this class. In fact, several variables of this type have been firstly discovered by their X-ray emission. This is the case of RBS~0324, which was identified as a polar candidate in the course of the ROSAT Bright Survey (RBS) (Schwope et al. 2000). For reviews on polars and cataclysmic variables, see Cropper (1990), Hellier (2001) and Warner (2005).

\vskip 10pt

The confirmation that RBS~0324 is really a polar was done by Schwope et al. (2002). The object shows broad, asymmetric emission lines of the Balmer series, HeI and HeII. They are typical of polars in the state of high accretion rate. Photometric and polarimetric observations revealed variations of brightness with timescales of minutes to hours and amplitude of 2 magnitudes, as well circular polarisation reaching 20$\%$ in certain phases, all modulated with a period of 181.8 min, which was considered as the orbital period of the binary.

\vskip 10pt

The fact that RBS~0324 is a poorly known polar motivated us to include it in our photometric, polarimetric and spectroscopic observational programs whose objective is better characterize magnetic cataclysmic variables and/or candidates to this class. In this letter, we present spectroscopic observations of this object.

\vskip 10pt

Spectra of RBS~0324 were collected on 2009, September 26 (UT) using the Goodman spectrograph at the 4.1-m SOAR telescope, Chile. We used the 1200 l/mm transmission VPH grating and a 0.84 arcsec long slit. The detector was a Fairchild 4096$\times$4096 CCD, with 15 micron/pixel (0.15 arcsec/pixel). This instrumental setup provided a spectral coverage from 4280 to 5580 \r{A} with a reciprocal dispersion of 0.31 \r{A}/pixel and a spectral resolution of 1.5 \r{A}. A total of 12 spectra was collected using an integration time of 15~minutes. The data cover more than one cycle of the 181.8 min orbital period. The reduction was done in the usual manner using the IRAF package\footnote{IRAF is distributed by the National Optical Astronomy Observatories, which are operated by the Association of Universities for Research in Astronomy, Inc., under cooperative agreement with the National Science Foundation.}, and consisted of zero and flat corrections, extraction to unidimensional spectrum, and finally wavelength and flux calibrations.

\vskip 10pt

The average spectrum of RBS~0324 is shown in Figure~1. Emission lines of Balmer series, HeI and HeII are very prominent, with HeII~$\lambda$4686 so intense as H$\beta$. This is usually seen in others magnetic cataclysmic variables and is used as one of the criteria for select candidates to this class. The individual spectra around the H$\gamma$, HeII~$\lambda$4686 and H$\beta$ spectral regions are shown in Figure~2. They are ordinated in phase using the ephemeris given by Schwope et al. (2002): HJD~$=$~2452262.0~$+$~0.126245~$\times$~E. The profiles are complex, presenting broad and narrow components, and are highly variable with the orbital phase. The profiles for the three lines are similar, this is particularly evident for HeII~$\lambda$4686 and H$\beta$ lines whose signal-to-noise ratios are better.

\vskip 10pt

We have determined the radial velocity of the H$\beta$ and HeII~$\lambda$4686 lines. Initially, each spectrum was continuum subtracted and smoothed due to the oversampling of the spectral PSF. To have an estimate of the radial velocity of each line, we calculated the flux weighted centroid as well as the line peak velocity. The resulting centroid radial velocity curves are shown in Figure~3 (top panel), where we used the same ephemeris as above. The two curves have the same pattern, with a semi-amplitude of about 250 km/s. An estimate of the uncertainty in the velocities was done by checking the dispersion of the sky line in $\lambda$5577. This provides an error of $\sim$20$-$30 km/s for the individual velocities. A sinusoid fit to the H$\beta$ narrow peak velocities suggests that the inferior conjunction of the secondary star occurs around phase 0.92+/-0.10. However, better S/N data is required to confirm the absolute phasing for this system.

\vskip 10pt

Figure~3 (bottom panel) shows the continuum variation along the orbital cycle. The continuum magnitude was estimated using a square band of 660 \r{A} width centred at 5230~\r{A}. We performed flux calibration using spectrophotometric standard stars (Hamuy et al. 1994). However, we do not consider the absolute calibration trustful due to significant slit losses. Even so, there is a clear modulation with orbital period with an amplitude of about 1.5~mag. This amplitude is compatible with that presented by Schwope et al. (2002) and shows its maximum around phase 0.55. The light and polarization curves of RBS~0324 (Schwope et al. 2002) are typical of cyclotron emission in polars. Therefore the origin of the modulation shown in Figure~3 should be cyclotron. However, around phase 0.55 the observer sees the illuminated surface of the secondary, which can also contribute to the flux modulation.

\vskip 10pt

These data represent the first approach on time-resolved spectroscopy of RBS~0324. The very complex line emission profiles put this object as a potential target for future and more detailed spectroscopic studies, such as Doppler tomography. These studies are important to understand the accretion structure in magnetic compact binaries.

\vskip 10pt

{\it Aknowledgements}. We thank the SOAR staff, particularly Luciano Fraga and S\'{e}rgio Scarano, for support during the observations. C.V. Rodrigues and K.M.G. Silva aknowledge CNPq and FAPESP grants, Processes 308005/2009$-$0 and 2008/09619$-$5, respectively. M.P. Diaz thanks the support under CNPq grant 305725.

\references

Schwope, A.D., Hasinger, G., Lehmann, I. et al., 2000, {\it AN}, {\bf 321}, 1 \BIBCODE{2000AN....321....1S}

Schwope, A.D., Brunner, H., Buckley, D., Greiner, J., v.d. Heyden, K., Neizvestny, S., Potter, S., Schwarz, R., 2002, {\it A\&A}, {\bf 396}, 895 \BIBCODE{ 2002A&A...396..895S}

Cropper, M., The Polars, 1990, {\it Space Science Reviews}, {\bf 54}, 195 \BIBCODE{1990SSRv...54..195C}

Hamuy, M., Suntzeff, N. B., Heathcote, S. R., Walker, A. R., Gigoux, P., Phillips, M. M., 1994, {\it PASP}, {\bf 106}, 566 \BIBCODE{ 1994PASP..106..566H}

Hellier, C., 2001, {\it Cataclysmic variable stars}, Chichester: Spring \BIBCODE{2001cvs..book.....H}

Warner, B., 1995, {\it Cataclysmic variable stars}, Cambridge University Press, Cambridge. \BIBCODE{1995CAS....28.....W}

\endreferences

\vskip 10pt

\IBVSfig{6.8cm}{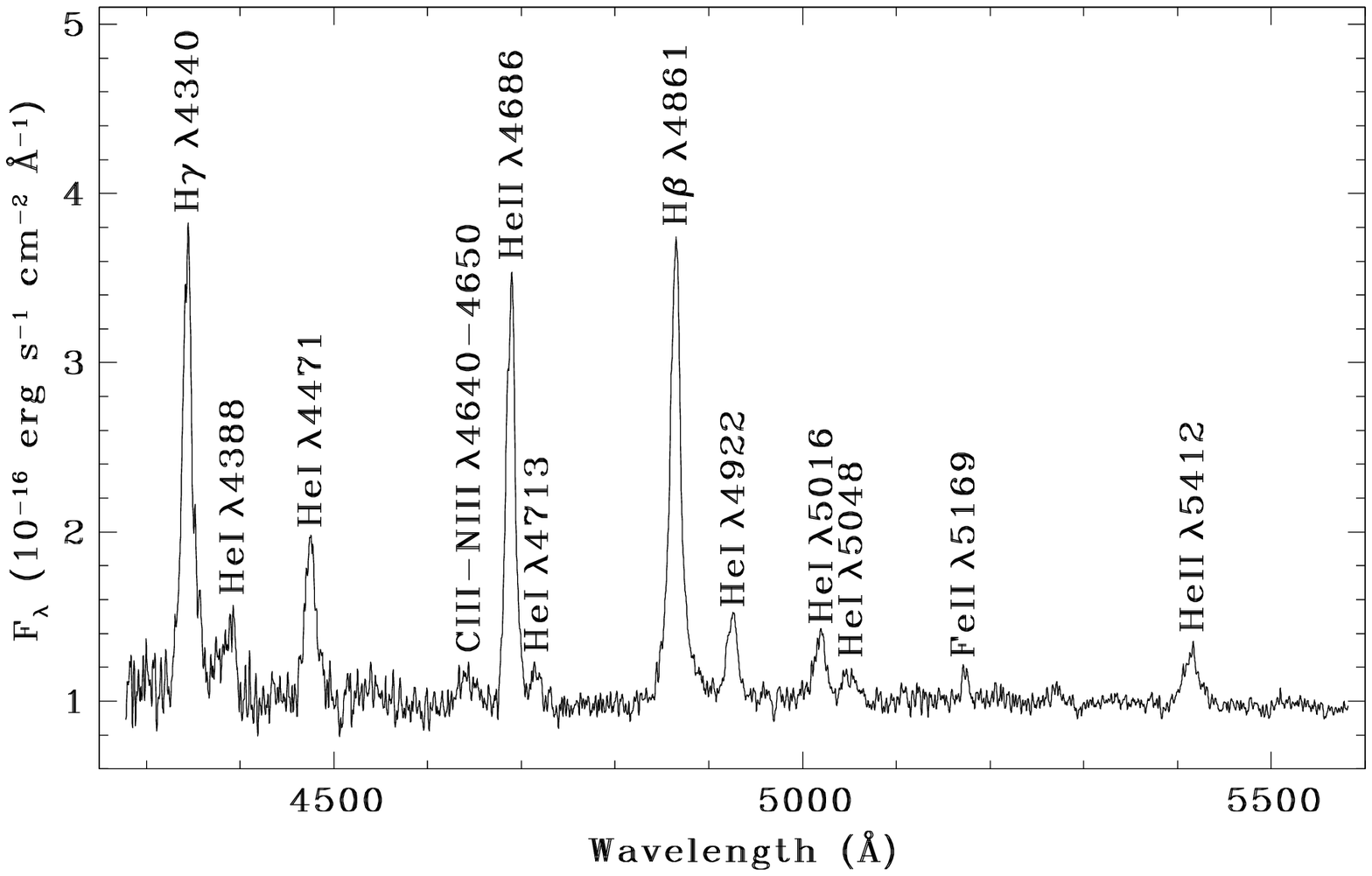}{Average spectrum of RBS~0324.}
\IBVSfigKey{fig_1.eps}{RBS~0324}{average spectrum}

\vskip 10pt

\IBVSfig{15.5cm}{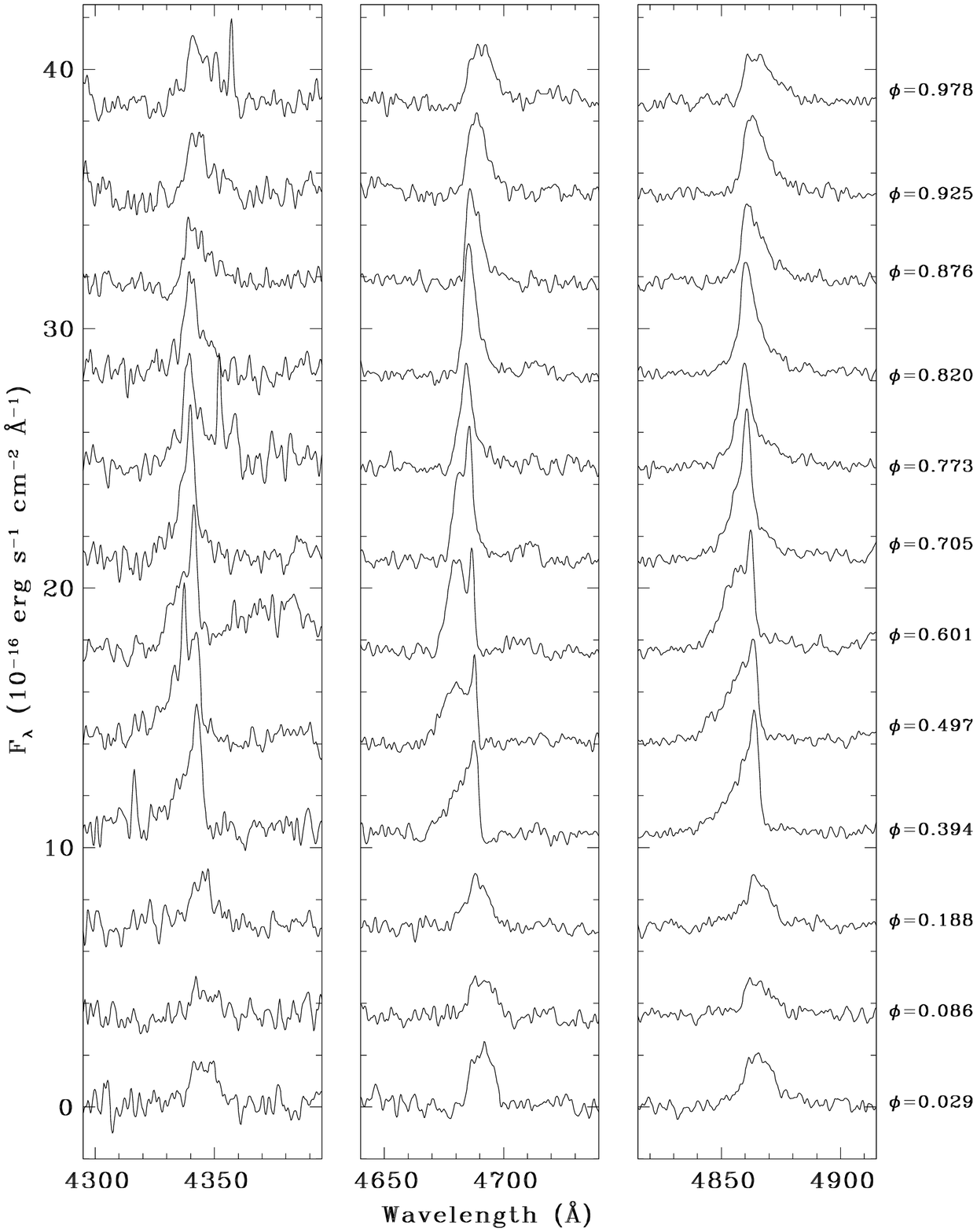}{H$\gamma$ (left), HeII~$\lambda$4686 (middle) and H$\beta$ (right) profiles as a function of the orbital phase of RBS~0324. The phases, $\phi$, were calculated using the ephemeris HJD~$=$~2452262.0~$+$~0.126245~$\times$~E (Schwope et al. 2002). The individual spectra were continuum subtracted and are arbitrarily shifted for better visualisation. }
\IBVSfigKey{fig_2.eps}{RBS~0324}{profiles}

\vskip 10pt

\IBVSfig{13.7cm}{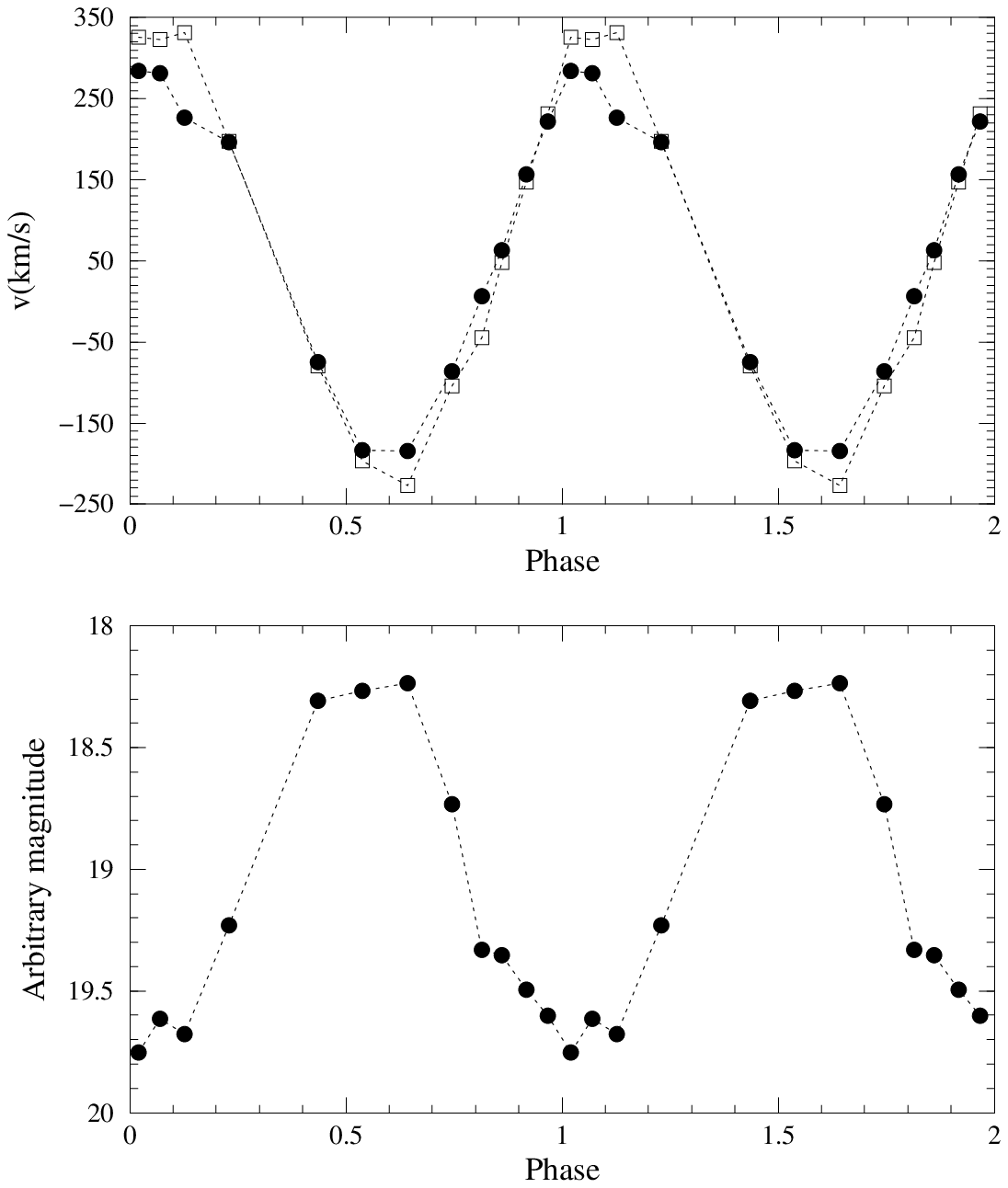}{The top panel shows the phase diagram of the radial velocities of RBS~0324 using H$\beta$ (filled circle) and HeII~$\lambda$4686 (open square). The bottom panel shows the phase diagram of the continuum in the interval from 4900 to 5560 \r{A}. The phases were calculated using the ephemeris provided by Schwope et al. (2002).}
\IBVSfigKey{fig_3.eps}{RBS~0324}{radial velocities}

\end{document}